\documentclass[letterpaper, 10 pt, conference]{ieeeconf}  % Comment this line out if you need a4paper

\IEEEoverridecommandlockouts                              % This command is only needed if
\usepackage{amsfonts,amssymb,mathrsfs,graphicx,amsmath,balance,color}

\usepackage{amsthm}
\newtheoremstyle{mystyle} % style name
            {3pt}         % vertical space above
            {3pt}         % vertical space below
            {\normalfont} % body font
            {}            % indent dim
            {\bfseries}   % Theorem head font
            {:}           % punctuation after theorem head
            {0.5em}       % space after theorem head
            {}            % theorem spec
\theoremstyle{mystyle}

\newtheorem{myRema}{\textit{Remark}}

\title{\LARGE \bf
Parallel Optimal Control for Cooperative Automation of Large-scale Connected Vehicles via ADMM
 }

\author{Zhitao Wang, Yang Zheng, Shengbo Eben Li*, Keyou You, and Keqiang Li% <-this % stops a space
\thanks{This study is supported by NSF China with 51575293 and 51622504, National Key R\&D Program of China with 2016YFB0100906, and International Sci\&Tech Cooperation Program of China under 2016YFE0102200. All correspondence should be sent to S. Li.}% <-this % stops a space
\thanks{Yang Zheng was with the Department of Automotive Engineering, Tsinghua University,
 Beijing, China. He is now with the Department of Engineering Science, University of Oxford, UK. ({\tt\small zhengy093@gmail.com}).}%
\thanks{Keyou You is with the Department of Automation, Tsinghua University, Beijing, China. ({\tt\small youky@tsinghua.edu.cn}).}%
\thanks{The other authors are with the Department of Automotive Engineering, Tsinghua University, Beijing, China. ({\tt\small wangzt16@mails.tsinghua.edu.cn, \{lishbo, likq\}@tsinghua.edu.cn}).}%
}

\begin{document}

\maketitle
\thispagestyle{empty}
\pagestyle{empty}

%%%%%%%%%%%%%%%%%%%%%%%%%%%%%%%%%%%%%%%%%%%%%%%%%%%%%%%%%%%%%%%%%%%%%%%%%%%%%%%%
\begin{abstract}

This paper proposes a parallel optimization algorithm for cooperative automation of large-scale connected vehicles. The task of cooperative automation is formulated as a centralized optimization problem taking the whole decision space of all vehicles into account. Considering the uncertainty of the environment, the problem is solved in a receding horizon fashion. Then, we employ the alternating direction method of multipliers (ADMM) to solve the centralized optimization in a parallel way, which scales more favorably to large-scale instances. Also, Taylor series is used to linearize nonconvex constraints caused by coupling collision avoidance constraints among interactive vehicles. Simulations with two typical traffic scenes for multiple vehicles demonstrate the effectiveness and efficiency of our method.

\end{abstract}
%%%%%%%%%%%%%%%%%%%%%%%%%%%%%%%%%%%%%%%%%%%%%%%%%%%%%%%%%%%%%%%%%%%%%%%%%%%%%%%%


%%%%%%%%%%%%%%%%%%%%%%%%%%%%%%%%%%%%%%%%%%%%%%%%%%%%%%%%%%%%%%%%%%%%%%%%%%%%%%%%
\section{Introduction}

With the rapid development of communication, computation, and automation technologies, connected vehicles become one key component in future transportation systems. Cooperative automation of connected vehicles is expected to be realized using powerful computing ability (\emph{e.g.}, cloud computation) and advanced communication technology (\emph{e.g.}, V2V, V2I)~\cite{6823640}. In recent years, coordinating multiple connected vehicles has received considerable research attention due to its potential to improve traffic efficiency and safety \cite{6121907,li2017dynamical}. One main challenge in cooperative automation is to ensure that coupling constraints among connected vehicles are satisfied and the computation is efficient~\cite{li2017simultaneous,kuwata2011cooperative}.

To coordinate multiple connected vehicles, one natural and promising approach is to use a distributed framework where each vehicle optimizes its own control variable based on local information. Recently, some distributed control laws for platooning of connected vehicles have been proposed, which employ matrix decomposition to decouple coupling constraints in the system. For example, Zheng \emph{et al}. introduced a distributed model predictive controller for heterogeneous platoons with unidirectional topologies in which closed-loop stability and robustness were analyzed~\cite{zheng2017distributed}. Wu \emph{et al}. proposed a distributed sliding mode algorithm based on a topologically structured function~\cite{wu2016distributed}. To deal with coupling constraints in planning problems, an effective solution is to use the priority approach, which has been widely applied in the literature~\cite{velagapudi2010decentralized}. In principle, planning problems are solved sequentially according to some prioritization to decouple the coupling constraints, thus achieving a distributed solution. For example, Kuwata \emph{et al}. introduced a cooperative distributed robust safe optimization approach for systems with coupled objectives and constraints, where the optimization was carried out in a sequential way~\cite{kuwata2011cooperative}.  Sycara \emph{et al}. applied a sparse method to improve the convergence performance, where constraints were discovered probabilistically~\cite{velagapudi2010decentralized}. Other techniques, such as graph searching, are also proposed for solving this problem \cite{plessen2016multi}. LaValle \emph{et al}. established a solution mapping between the planning problem and network-flow, and then applied integer linear programming to optimize the objective \cite{yu2016optimal}. In short, to deal with coupling constraints, the distributed formation of the problem needs to be assigned appropriately, \emph{i.e.}, the topology of platoons and the priorities of the priority approach \cite{li2017dynamical,velagapudi2010decentralized,zheng2016stability}.

Another intuitive way of executing a cooperative task is to establish a centralized system to optimize the global decision space of all vehicles. However, the centralized formulation is computationally expensive when the system size increases~\cite{li2017simultaneous}. One way to mitigate the scalability issue is to develop powerful numerical solvers to handle large-scale optimization problems~\cite{richards2002aircraft,borrelli2006comparison,zheng2017chordal}. For example, Borrelli~\emph{et al}. used IPOPT to solve optimization problems in trajectory planning tasks of multiple UAVs~\cite{borrelli2006comparison}. A scalable first-order solver has recently been developed for sparse conic problems~\cite{zheng2017chordal}. These solvers could handle the networks of small size efficiently, but may still scale poorly to large-scale and dense instances. Some researches take advantage of special structures of multi-agent formations to improve computation performance. For instance, Alonso-Mora \emph{et al}. studied the formation of multiple vehicles and object transport by dividing the task into obstacle-free region computation and formation parameters optimization~\cite{alonso2017multi}. Urcola \emph{et al}. proposed a hybrid centralized-distributed architecture for the path planning for robot formations, where the leader centralizes the information and executes the global process and the followers execute the navigation in a distributed way~\cite{urcola2017cooperative}. Zheng \emph{et al.} introduced an efficient sequential algorithm for scalable design of structured controllers by exploiting underlying sparsity properties~\cite{zheng2018scalable}. While efficient computation could be obtained through these approaches, the special nature of the tasks prevents the application of the algorithms to general connected vehicle networks.

In this paper, we formulate the cooperative automation of multiple connected vehicles as a centralized optimal control problem. The receding horizon framework is applied to ensure its robustness in dynamic environment and disturbances. To extend the problem solution to large-scale cases, we use the alternating direction method of multipliers (ADMM)~\cite{boyd2011distributed} to decompose the centralized optimization problem. One notable feature is that parallel computation can be realized after the decomposition using ADMM. Taylor series is utilized to linearize the nonconvex constraints in the problem to facilitate the application of ADMM. Note that ADMM is a simple yet powerful algorithm for distributed optimization, which has been applied in a wide range of fields \cite{boyd2011distributed,bento2013message}. The contributions of this paper are 1) we design an optimal control framework for coordinating connected vehicles, in which the cooperative automation task is formulated as a centralized optimization problem in a receding horizon fashion; 2) we propose a parallel optimization algorithm to solve the centralized optimization problem via ADMM, and the algorithm can be implemented on a cloud platform to deal with large-scale vehicle networks.

The rest of this paper is organized as follows. Section II formulates the centralized optimal control problem for the cooperative automation of connected vehicles; Section III introduces the parallel computation algorithm via ADMM; Section IV demonstrates the effectiveness of the proposed algorithm, and we conclude this paper in Section V.

%%%%%%%%%%%%%%%%%%%%%%%%%%%%%%%%%%%%%%%%%%%%%%%%%%%%%%%%%%%%%%%%%%%%%%%%%%%%%%%%%%%%%%%
%%%%%%%%%%%%%%%%%%%%%%%%%%%%%%%%%%%%%%%%%%%%%%%%%%%%%%%%%%%%%%%%%%%%%%%%%%%%%%%%%%%%%%%

\section{Cooperative Automation of Connected Vehicles}\label{section II}

Connected vehicles equipped with advanced sensing and communication devices can sense the surroundings and upload information to a cloud platform. Then, the cloud platform computes the decision action and transmits instructions to each vehicle for coordination.

%%%%%%%%%%%%%%%%%%%%%%%%%%%%%%%%%%%%%%%%%%%%%%%%%%%%%%%%%%%%%%%%%%%%%%%%%%%%%%%%%%%%%%%

\subsection{Modelling of Computation Networks}

An undirected graph $\mathcal{G}(\mathcal{V}, \mathcal{E})$ is used to describe the constraint topology among interactive vehicles, where $\mathcal{V} = \{1,...,N\}$ denotes a set of connected vehicles in the system and $N$ is the number of vehicles. Since only the vehicles in the proximity of another one have the collision possibility, it is not necessary to consider the constraints with all the other vehicles. $\mathcal{E} = \{1,...,M\}$ is the set of edges which represent the coupling constraint between two interactive vehicles, defined as
\begin{equation}
\begin{cases}
(i,j)\in\mathcal{E}(t)\quad \Vert{p_i-p_j}\Vert\leqslant d_{\rm perc} \\
(i,j)\notin\mathcal{E}(t)\quad \rm otherwise,
\end{cases}
\end{equation}
where $M$ is the number of coupling constraints in the vehicle network, $p_i = [p_{i,x}, p_{i,y}]^T$ is position of vehicle $i$ in a global coordinate system, $d_{\rm perc}$ is the distance which ensures collision avoidance between two vehicles under the constraint of vehicle kinematics. The perception distance of each vehicle should be not less than $d_{\rm perc}$, which guarantees that the vehicles linked by $(i,j)$ could take action in time as soon as sensing the information of the coupling vehicles. We define the neighbor nodes of each vehicle as
\begin{equation}
\mathcal{N}_i=\{j|(i,j)\in\mathcal{E},\forall j\in\mathcal{V}\}.
\end{equation}

\begin{figure}[t]%[hb]
\centering
  \setlength{\belowcaptionskip}{0em}
\includegraphics[scale=0.42]{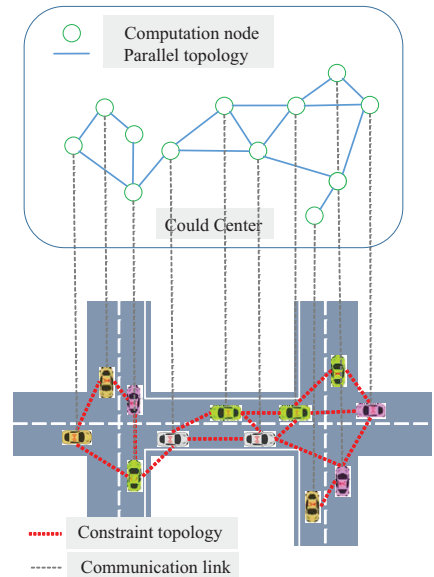}
\caption{An illustration of vehicle network and computation network of the cloud platform.}
\label{figurelabel1}
\end{figure}

In addition to the constraint topology among vehicles, the computation network of the cloud platform can be described by $\mathcal{G}$ as well. The cloud platform is composed of computer clusters which allows high-performance parallel computation. The computation nodes are the mapping of vehicle nodes in the constraint topology. The relation between the constraint topology and the computation network is shown in Fig.~\ref{figurelabel1}.

\begin{myRema}
Since the vehicles move over time, $\mathcal{G}$ is time-varying and the links between computation nodes $\mathcal{V}$ change as well. To ensure efficient computation, the link and the information transmission between computation nodes should be managed by an effective coordination mechanism. The computation resource scheduling and management are beyond the scope of this paper.
\end{myRema}

%%%%%%%%%%%%%%%%%%%%%%%%%%%%%%%%%%%%%%%%%%%%%%%%%%%%%%%%%%%%%%%%%%%%%%%%%%%%%%%%%%%%%%%

\subsection{A Centralized Large-scale Optimal Control Problem}

Each connected vehicle in $\mathcal{V}$ has its own origin and target position, which needs to finish its own guidance task in a coordinating fashion with other vehicles. The path of each vehicle can be planned by any efficient planning algorithms, \emph{e.g.}, RRT and A* algorithm \cite{paden2016survey}. Then, the cooperative automation task can be formulated as an optimization problem, in which each vehicle is scheduled to plan its path and follow the planned path under the constraints of physical environments, \emph{i.e.}, collision avoidance constraints and other physical limitations.

In this paper, we formulate the automation problem in a centralized fashion based on the receding horizon optimization framework as
\begin{equation}\label{equ:centr_pro}
\begin{aligned}
 \min_{u_i} \quad & J(u_i)=\sum_{i\in\mathcal{V}}\Big[\int_t^{t+T} h_i(x_i,u_i)d\tau\Big]\\
 \text{subject to} \quad & \dot{x}_i=f_i(x_i,u_i)\\
 & (x_i,u_i)\in\mathbb{C}_{\rm s}\\
 & (x_i,u_i,x_j,u_j)\in\mathbb{C}_{\rm c}, j\in\mathcal{N}_i,
\end{aligned}
\end{equation}
where $x_i=[p_i,{\theta}_i]$ and $u_i$ denote the state and control variable of vehicle $i$, respectively; $\dot{x}_i=f(x_i,u_i)$ denotes the kinematics of the vehicle; $T$ is the predicted time horizon; $\mathbb{C}_{\rm s}$ denotes the set of constraints of the vehicle $i$; and $\mathbb{C}_{\rm c}$ is the set of coupling constraints of a pairwise of connected vehicles $(i,j), j\in\mathcal{N}_i$, the function $h_i(x_i,u_i)=Q\Vert{x_i-x_{\rm ref,\textit{i}}}\Vert_2^2$ describes the deviation of the predicted planning trajectory from the reference trajectory $x_{\rm ref,\textit{i}}$.

Specifically, $\mathbb{C}_{\rm s}$ represents the constraints of the physical environment for a single vehicle, \emph{e.g.}, the bound of the steering angle and the vehicle position. These can be described as
\begin{equation}\label{equ:C_s}
\mathbb{C}_{\rm s}=\{(x,u)|x_{\rm low}\leqslant x \leqslant x_{\rm up}, u_{\rm low}\leqslant u\leqslant u_{\rm up}\},
\end{equation}
where $\mathbb{C}_{\rm c}$ represents the collision avoidance constraints between pairwise interactive vehicles. Based on the constraint topology, this is imposed as
\begin{equation}\label{equ:C_c}
\mathbb{C}_{\rm c}=\{(x_i,u_i,x_j,u_j)|\Vert {p_i-p_j}\Vert\geqslant d_{\rm safe}, j\in\mathcal{N}_i\},
\end{equation}
where $d_{\rm safe}$ is a predefined safe separation distance between vehicles.

%%%%%%%%%%%%%%%%%%%%%%%%%%%%%%%%%%%%%%%%%%%%%%%%%%%%%%%%%%%%%%%%%%%%%%%%%%%%%%%%%%%%%%%%

\subsection{The Receding Horizon Optimization}

To cope with the dynamic environment and disturbance, receding horizon optimization is carried out to solve the optimal control problem. It solves the optimization problem (\ref{equ:centr_pro}) to generate control inputs for a finite horizon starting from the latest states, where only the first step of the control inputs is implemented.

It is assumed that the network $\mathcal{G}$ is time-invariant in the predicted time horizon $\tau\in[t,t+T]$ for convenience of computations, \emph{i.e.}, $\mathcal{G}(\tau)=\mathcal{G}(t)$. This assumption makes the coupling constraints time-invariant. The selection of predicted time horizon $T$ should consider the kinematic of the bicycle model and ensure the pairwise vehicles keep safe distance once they sense each other. Every vehicle in the centralized formulation is coupled due to the pairwise collision
\begin{figure}[t]%[ht]
\centering
\includegraphics[scale=0.4]{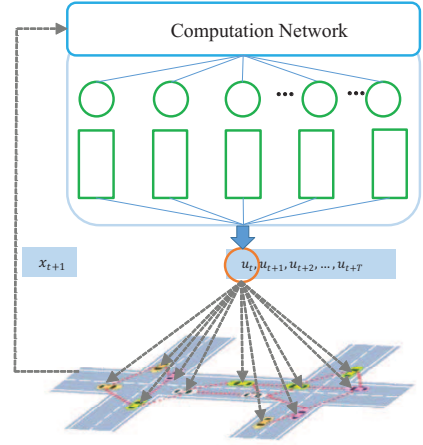}
\caption{The receding horizon optimization framework.}
\label{figurelabel2}
\end{figure}
avoidance constraints. A decentralized solution is proposed to solve this centralized formulation in the following section.

\begin{myRema}
The centralized problem formulation makes it easy to ensure the stability of model predictive control compared with Kuwata's work \cite{kuwata2011cooperative}. Also, this formulation does not need to design any coordination mechanism compared to the priority approach and related distributed control \cite{velagapudi2010decentralized}.
\end{myRema}

%%%%%%%%%%%%%%%%%%%%%%%%%%%%%%%%%%%%%%%%%%%%%%%%%%%%%%%%%%%%%%%%%%%%%%%%%%%%%%%%%%%%%%
%%%%%%%%%%%%%%%%%%%%%%%%%%%%%%%%%%%%%%%%%%%%%%%%%%%%%%%%%%%%%%%%%%%%%%%%%%%%%%%%%%%%%%%%

\section{Parrallel Optimization Algorithm}

The centralized optimal control \eqref{equ:centr_pro} can be reformulated as a quadratic programming (QP) problem in a discrete time domain. While interior-point methods are efficient to solve~\eqref{equ:centr_pro} when the problem is of small size, the computational demand increases as the number of vehicles grows, which scales poorly to large-scale instances. Decentralized optimization algorithms based on first-order algorithms that scale well to large-scale instances have attracted considerable research attention in recent years \cite{boyd2011distributed}. ADMM is one efficient first-order algorithm to distribute the computation of a large-scale problem to a network of computing nodes, which has been applied in many fields, \emph{e.g.}, statistical learning \cite{boyd2011distributed}, distributed control \cite{ong2015cooperative}, power system management \cite{erseghe2014distributed} and sparse semidefinite programs \cite{zheng2017fast}. Motivated by the wide applications, we employ ADMM to solve our large-scale connected automation problem in a parallel way.

In this section, we first transform the centralized optimization problem into a consensus optimization problem. Then, ADMM is applied to decompose the consensus problem, and parallel computation can be realized based on the decomposition through the cloud computation network.

%%%%%%%%%%%%%%%%%%%%%%%%%%%%%%%%%%%%%%%%%%%%%%%%%%%%%%%%%%%%%%%%%%%%%%%%%%%%%%%%%%%%%

\subsection{The ADMM Algorithm}

The ADMM algorithm aims to solve an optimization problem in the following form~\cite{boyd2011distributed}
$$
\begin{aligned}
 \min \quad & F(a)+G(b)\\
\text{subject to} \quad & Aa+Bb=d,\\
\end{aligned}
$$
where $F$ and $G$ are convex function, $a\in\mathbb{R}^{n_a},b\in\mathbb{R}^{n_b},d\in\mathbb{R}^{n_d}, A\in\mathbb{R}^{n_d \times n_a}$ and $B\in\mathbb{R}^{n_d \times n_b}$. Given a penalty parameter $\rho > 0$ and a scaled dual multiplier $z\in\mathbb{R}^{n_d}$, the (scaled) ADMM algorithm minimizes the augmented Lagrangian
$$
L_{\rho}(a,b,z)=F(a)+G(b)+\frac{\rho}{2}\Vert{Aa+Bb-d+z}\Vert^2,
$$
with respect to the variables $a$ and $b$ separately, followed by a dual variable update:
\begin{equation}\label{equ:ADMM}
\begin{aligned}
a^{k+1} & :=\underset{a}{\text{argmin}}\, L_{\rho}(a,b^{k},z^{k}),\\
b^{k+1} & :=\underset{b}{\text{argmin}}\, L_{\rho}(a^{k+1},b,z^{k}),\\
z^{k+1} & := z^k+(Aa^{k+1}+Bb^{k+1}-d).
\end{aligned}
\end{equation}
The superscript $k$ indicates that a variable is fixed to its value at the $k$-th iteration. ADMM is particularly suitable when the minimizations with respect to each of the variables $a$ and $b$ in \eqref{equ:ADMM} can be carried out efficiently. We refer the interested readers to~\cite{boyd2011distributed} for more details.

%%%%%%%%%%%%%%%%%%%%%%%%%%%%%%%%%%%%%%%%%%%%%%%%%%%%%%%%%%%%%%%%%%%%%%%%%%%%%%%%%%%%%%%

\subsection{Formulation of the Consensus Optimization}

To deal with $\mathbb{C}_{\rm s}$ and $\mathbb{C}_{\rm c}$ in (\ref{equ:centr_pro}), two indicator functions $\mathbb{I}_{\rm s}(\cdot)$ and $\mathbb{I}_{\rm c}(\cdot)$ are defined to describe the constraints of the vehicle $i$ and pairwise interactive vehicles $(i,j)$ respectively:
\begin{align}\label{equ:Indicator}
&\mathbb{I}_{\rm c}(x_i,u_i,x_j,u_j):= \notag\\
&\begin{cases}
0\quad (x_i,u_i,x_j,u_j)\in \mathbb{C}_{\rm c}, j\in \mathcal{N}_i\text{ and } \dot{x}_i=f_i(x_i,u_i) \\
\infty\quad \rm otherwise,
\end{cases} \notag\\
&\mathbb{I}_{\rm s}(x_i,u_i):= \notag\\
&\begin{cases}
0\quad (x_i,u_i)\in \mathbb{C}_{\rm s}\text{ and } \dot{x}_i=f_i(x_i,u_i) \\
\infty\quad \rm otherwise.
\end{cases} %\\
\end{align}
%\end{equation}
%
Then, the centralized problem \eqref{equ:centr_pro} can be equivalently rewritten as
\begin{equation}\label{equ:newCenPro}
\begin{aligned}
\min_{u_i} \; \hat{J}(u_i)=&\sum_{i\in\mathcal{V}}\Big[\int_t^{t+T}h_i(x_i,u_i)d\tau+\mathbb{I}_{\rm s}(x_i,u_i)\Big]+\\
&\sum_{(i,j)\in\mathcal{E}}\mathbb{I}_{\rm c}(x_i,u_i,x_j,u_j).
\end{aligned}
\end{equation}

The optimization problem (\ref{equ:newCenPro}) is coupled among $\mathcal{V}$ due to the coupling constraint $\mathbb{I}_{\rm c}(\cdot)$. By introducing a set of consensus constraints, problem (\ref{equ:newCenPro}) could be equivalently rewritten into the standard ADMM form
\begin{equation}\label{equ:stanADMM}
\begin{aligned}
\min_{u_v,u_v^e}
& \sum_{v\in\mathcal{V}}\Big[\int_t^{t+T}h_v(x_v,u_v)d\tau+\mathbb{I}_{\rm s}(x_v,u_v)\Big]+\\
&\sum_{e\in\mathcal{E}}\mathbb{I}_{\rm c}(\{x_v^e,u_v^e\}_{v\in\mathcal{V}(e)})\\
\text{Subject to:}
& \quad u_v=z_v, \; u_v^e=z_v,\\
& \quad v\in\mathcal{V},e\in\mathcal{E}(v),
\end{aligned}
\end{equation}
where $\mathcal{V}(e)$ represents the set of computation nodes linked with edge $e\in\mathcal{E}$, which called local nodes, $\mathcal{E}(v)$ represents the set of link nodes linked with node $v\in\mathcal{V}$, and $\{x_v^e,u_v^e\}_{v\in\mathcal{V}(e)}$ denotes a vector indexed by a parameter $v\in\mathcal{V}(e)$, $z_v$ is the consensus variable.

\begin{figure}[t]
\centering
\includegraphics[scale=0.42]{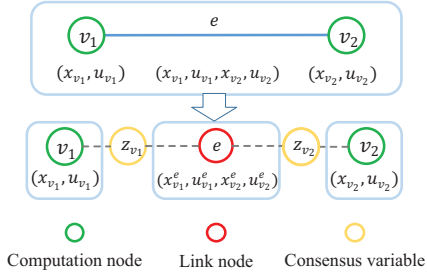}
\caption{Formulation of the consensus optimization: consensus constraints are applied to decompose the coupling.}
\label{figurelabel3}
\vspace{-2mm}
\end{figure}

For the purpose of parallel computation, we assign link node to each pairwise interactive local node and arrange the single vehicle objective function and coupled objective function in (\ref{equ:stanADMM}) to local and link node respectively with consensus constraint exist during local and link nodes. Next, the parallel algorithm will be realized based on ADMM. Fig.~\ref{figurelabel3} illustrates the formulation of the consensus optimization.

%%%%%%%%%%%%%%%%%%%%%%%%%%%%%%%%%%%%%%%%%%%%%%%%%%%%%%%%%%%%%%%%%%%%%%%%%%%%%%%%%%%%%%%%

\subsection{A Parallel Algorithm based on ADMM}

Here, we apply the ADMM to decompose the centralized consensus optimization problem (\ref{equ:stanADMM}), leading to a parallel algorithm to solve the problem. First, the augmented Lagrangian function for the problem (\ref{equ:stanADMM}) is
%\begin{equation}
\begin{align} \label{equ:myLagr}
& L_{\rho}(u_v,u_v^e,z_v,\lambda_v^e)=  \sum_{v\in\mathcal{V}}\Big[\int_t^{t+T}h_v(x_v,u_v)d\tau+\mathbb{I}_{\rm s}(x_v,u_v)\Big] \notag\\
& + \sum_{e\in\mathcal{E}}\mathbb{I}_{\rm c}(\{x_v^e,u_v^e\}_{v\in\mathcal{V}(e)})+\sum_{v\in\mathcal{V}(e)}\frac{\rho}{2}\Vert{u_v-z_v+\lambda_v}\Vert_2^2 \notag\\
& + \sum_{e\in\mathcal{E}}\sum_{v\in\mathcal{V}(e)}\frac{\rho}{2}\Vert{u_v^e-z_v+\lambda_v^e}\Vert_2^2,
\end{align}
%\end{equation}
where $\lambda_v$ and $\lambda_v^e$ are scaled dual variables corresponding to $u_v$ and $\{u_v^e\}_{v\in\mathcal{V}(e)}$ respectively, and $\rho$ is the augmented Lagrangian parameter.

According to (\ref{equ:ADMM}), the ADMM algorithm minimizes the augmented Lagrangian (\ref{equ:myLagr}) iteratively. At each iteration $k$, the algorithm consists of the following three steps.

\emph{Step 1}: Each local node $v\in\mathcal{V}$ solves the following local problem in parallel to update $u_v$:
\begin{align}\label{equ:Step1.1}
x_v^{k+1},u_v^{k+1}:=
\underset{u_v}{\text{argmin}}&\Big\{  \int_t^{t+T}h(x_v,u_v)d\tau+\mathbb{I}_{\rm s}(x_v,u_v)+ \notag\\
& \sum_{e\in\mathcal{E}(v)}\big[\frac{\rho}{2}\Vert{u_v-z_v^k+\lambda_v^k}\Vert_2^2\big] \Big\}.
\end{align}
Each link node solves the following coupled optimization problem locally in a parallel way to update $u_v^e$:
\begin{align}\label{equ:Step1.2}
\{x_v^e,u_v^e\}_{v\in\mathcal{V}(e)}^{k+1}:=
&\underset{u_v^e}{\text{argmin}}\Big\{  \mathbb{I}_{\rm c}(\{x_v^e,u_v^e\}_{v\in\mathcal{V}(e)})+ \notag\\
& \quad \sum_{v\in\mathcal{V}(e)}\big[\frac{\rho}{2}\Vert{u_v^e-z_v^k+{\lambda_v^e}^k}\Vert_2^2\big] \Big\}.
\end{align}

\begin{figure}[t]
\centering
\includegraphics[scale=0.4]{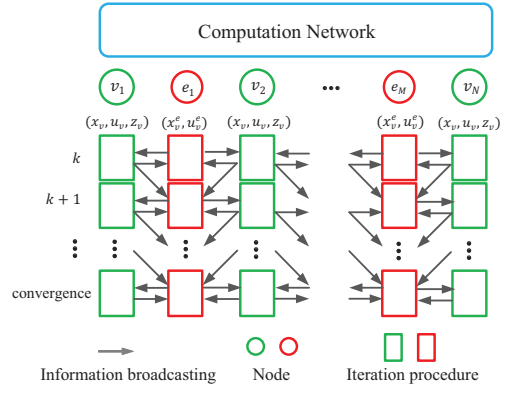}
\caption{An illustration of the parallel algorithm: the local and link nodes run their own task and change information with neighbors. }
\label{figurelabel4}
\vspace{-2mm}
\end{figure}

\emph{Step 2}: Each local node updates the consensus variable $z_v$ in the following form:
\begin{equation}\label{equ:Step2}
z_v^{k+1}:=\frac{\big(u_v^{k+1}+\frac{\lambda_v^k}{\rho}\big)+\sum_{e\in\mathcal{E}(v)}\big({u_v^e}^{k+1}+\frac{{\lambda_v^e}^k}{\rho}\big)}{1+\sum_{e\in\mathcal{E}(v)}1}.
\end{equation}

\emph{Step 3}: The scaled dual variable $\lambda$ is updated as (\ref{equ:Step3.1}) and (\ref{equ:Step3.2}), which could be computed on each local node and link node respectively:
\begin{align}
\lambda_v^{k+1}:= & \lambda_v^{k}+(u_v^{k+1}-z_v^{k+1}), \label{equ:Step3.1}\\
{\lambda_v^e}^{k+1}:= & {\lambda_v^e}^{k}+({u_v^e}^{k+1}-z_v^{k+1}). \label{equ:Step3.2}
\end{align}
The steps (\ref{equ:Step1.1})-(\ref{equ:Step3.2}) iterate on the computation nodes in parallel until convergence. In each step, the computation procedure is entirely decentralized to each node in the cloud network. By broadcasting information with neighbor computation nodes, the resulting algorithm can be carried out in a parallel way. Fig.~\ref{figurelabel4} shows the process and information broadcasting during each iteration of the ADMM algorithm.

\begin{myRema}
 Compared with \cite{ong2015cooperative,zheng2017fastadmm}, our algorithm is more suitable for parallel computation since the single and coupled objective functions are solved in parallel by introducing consensus variables rather than in sequence.
\end{myRema}

%%%%%%%%%%%%%%%%%%%%%%%%%%%%%%%%%%%%%%%%%%%%%%%%%%%%%%%%%%%%%%%%%%%%%%%%%%%%%%%%%%%%%%%

\subsection{Stopping Conditions}

According to \cite{boyd2011distributed}, the stopping criterion can be described with the primal and dual residuals as follows:
\begin{equation}\label{equ:StpCrt}
\begin{aligned}
r^{k+1}=&\Vert{u^{k+1}-z^{k+1}}\Vert_2\leqslant\epsilon^{\rm pri},\\
s^{k+1}=&\rho\Vert{z^{k+1}-z^{k}}\Vert_2\leqslant\epsilon^{\rm dual},
\end{aligned}
\end{equation}
where $\epsilon^{\rm pri}$ and $\epsilon^{\rm dual}$ are primal and dual feasibility tolerances respectively:
%\begin{equation}
\begin{align} \label{equ:PriAAbs}
\epsilon^{\rm pri}=&\epsilon^{\rm abs}\sqrt{(N+2M)N_p}+\epsilon^{\rm rel}\text{max}\{\Vert{u^k}\Vert_2,\Vert{z^k}\Vert_2\}, \notag\\
\epsilon^{\rm dual}=&\epsilon^{\rm abs}\sqrt{(N+2M)N_p}+\epsilon^{\rm rel}\frac{\Vert{\lambda^k}\Vert_2}{\rho},
\end{align}
%\end{equation}
where $\epsilon^{\rm abs}$ and $\epsilon^{\rm rel}$ are the absolute and relative criterion, $M$ is the size of link computation node group and $N_p$ is the number of predictive step.

While the ADMM algorithm converges independently of the choice of penalty parameter $\rho$, in practice this value strongly influences the number of iterations required for convergence. Unfortunately, analytic results for the optimal choice of $\rho$ are not available except for some special problems \cite{zheng2017chordal}. As suggested in \cite{boyd2011distributed}, in order to improve the practical convergence performance and make performance less dependent on the choice of $\rho$, our algorithm employs the following dynamic adaptive rule:
\begin{equation}
\rho^{k+1}=
\begin{cases}
\begin{aligned}
\tau^{\rm incr}\rho^{k}\quad &\Vert{r^k}\Vert_2>\mu\Vert{s^k}\Vert_2, \\
\frac{\rho^k}{\tau^{\rm decr}}\quad &\Vert{s^k}\Vert_2>\mu\Vert{r^k}\Vert_2,\\
\rho^k \quad &\text{otherwise},
\end{aligned}
\end{cases}
\end{equation}
where $\tau^{\rm incr}, \tau^{\rm decr}$ and $\mu$ are set to 2, 2 and 5 in our algorithm.

%%%%%%%%%%%%%%%%%%%%%%%%%%%%%%%%%%%%%%%%%%%%%%%%%%%%%%%%%%%%%%%%%%%%%%%%%%%%%%%%%%%%%%

\subsection{Convexification of the Problem}

The convergence of ADMM is typically guaranteed for convex problems~\cite{boyd2011distributed}. Here, we introduce a procedure to linearize the non-convex constraints due to the collision avoidance (\ref{equ:C_c}). Based on the properties of receding horizon optimization, we generate a seed trajectory $r^0$ by moving the optimal control input calculated in previous cycle one step forward and repeating the last prediction step~\cite{zheng2017fastadmm}.

\begin{figure}[t]%[ht]
\centering
\includegraphics[scale=0.4]{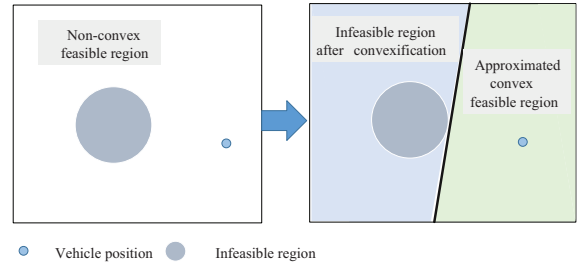}
\caption{Approximation of the nonconvex constraint: by linearization, a convex while sub-optimal decision space is got.}
\label{figurelabel5}
\end{figure}

To linearize the non-convex constraints, the Taylor's series is applied to the constraints \eqref{equ:C_c} at $p^0$:
\begin{equation}\label{equ:convex}
2(p_i^0-p_j^0)^T(p_i-p_j)-\Vert{p_i^0-p_j^0}\Vert_2^2\geqslant d_{\rm safe}^2, j\in \mathcal{N}_i.
\end{equation}
By linearization, we approximate the original nonconvex constraint with a convex linear constraint, which returns a sub-optimal planning path in the prediction time horizon for each vehicle in the network. Fig.~\ref{figurelabel5} illustrates the approximation of the nonconvex constraint.

\begin{figure}[t]%[ht]
\centering
\setlength{\abovecaptionskip}{0pt}
\includegraphics[scale=0.4]{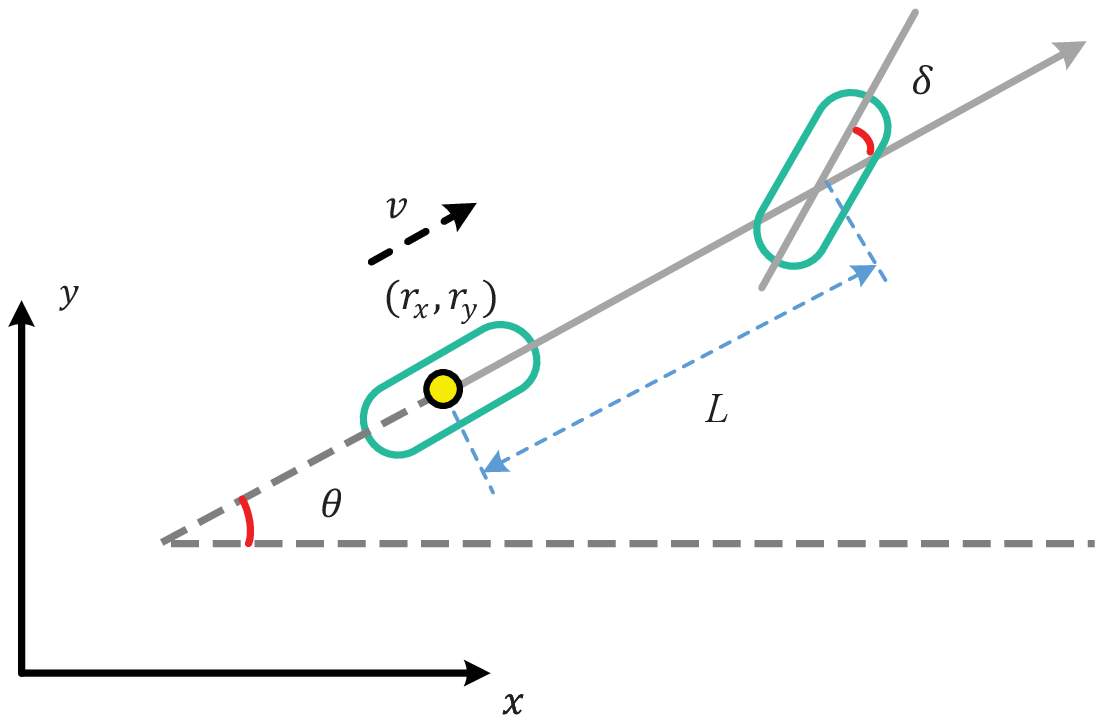}
\caption{The bicycle model.}
\label{figurelabel6}
\vspace{-2mm}
\end{figure}

%%%%%%%%%%%%%%%%%%%%%%%%%%%%%%%%%%%%%%%%%%%%%%%%%%%%%%%%%%%%%%%%%%%%%%%%%%%%%%%%%%%%%%%%
%%%%%%%%%%%%%%%%%%%%%%%%%%%%%%%%%%%%%%%%%%%%%%%%%%%%%%%%%%%%%%%%%%%%%%%%%%%%%%%%%%%%%%%%

\section{Numerical Simulations and Discussion}

In our simulation, the kinematics of each vehicle ignoring tire slip angle are described as follows:
\begin{equation}\label{equ:kine}
\begin{aligned}
\dot{r}_x = &v\cos\theta,\\
\dot{r}_y = &v\sin\theta,\\
\dot{\theta} = &\frac{v}{L}\tan\delta,
\end{aligned}
\end{equation}
where $(r_x,r_y)$ denotes the rear wheel axle center coordinates of the vehicle, $\theta$ denotes the vehicle heading angle with respect to the global $x$-axis (positive counter-clockwise), $v$ denotes the speed of rear wheel axle center, $\delta$ denotes the steer angle (positive counter-clockwise), and $L$ is the vehicle wheelbase. The position $(p_x,p_y)$ of the vehicle could be derived from these variables. See Fig.~\ref{figurelabel6} for an illustration.% of the bicycle model.

Note that the kinematic bicycle model, described as (\ref{equ:kine}), is non-linear and coupled between lateral and longitudinal kinematics. %Although the parallel algorithm is suitable to nonlinear state space model as well, but the computation for nonlinear problem is time consuming. 
To simplify the formulation and facilitate solution to the control problem, a common method is to assume that the vehicle speed is preset and constant in a short predictive horizon \cite{zheng2017fastadmm}. Then, we apply Taylor series to the kinematic model to transform the vehicle kinematics to a linear state space model~\cite{ong2015cooperative}.

\begin{figure}[t]%[ht]
\centering
\includegraphics[scale=0.34]{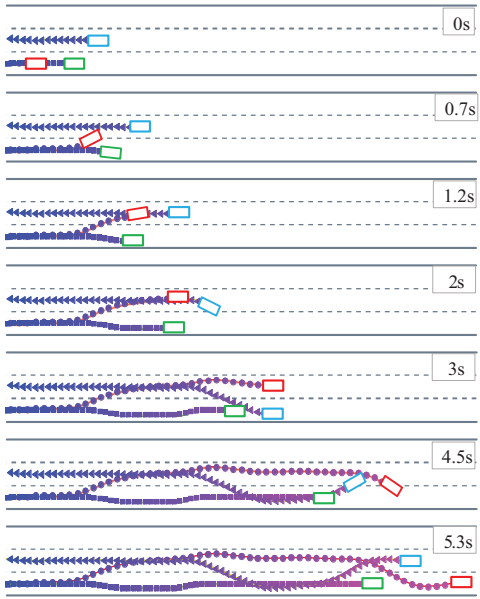}
\caption{Snapshots of cooperative overtaking of three vehicles.}
\label{figurelabel7}
\vspace{-2mm}
\end{figure}

Two typical traffic scenes with multiple connected vehicles are simulated to test the effectiveness of the proposed algorithm. The vehicle speeds are from 40km/h to 50km/h, and the wheelbase of vehicle is set as $L$=2.4m. Considering the tradeoff between the computation demanding and tracking performance, we set the sampling time $T_s$=0.1s and the prediction horizon $N_p$=15 in the simulation. The parameters of ADMM algorithm are set as following $\epsilon^{\rm abs}$=0.01, and $\epsilon^{\rm rel}$=0.01. The computation is implemented on a system with Intel i7 processor at 3.4 GHz. The ADMM algorithm is implemented in MATLAB environment with the quadratic programs solved by the interior-point solver Gurobi \cite{optimization2017inc}.

%%%%%%%%%%%%%%%%%%%%%%%%%%%%%%%%%%%%%%%%%%%%%%%%%%%%%%%%%%%%%%%%%%%%%%%%%%%%%%%%%%%%%%%

\subsection{Numerical Simulation}

The first traffic task is the cooperative overtaking of multiple vehicles in three lanes, a normal scenario in highway driving. In our simulation, three vehicles have different reference paths and desired velocities. The trajectories generated form the algorithm are shown in Fig.~\ref{figurelabel7}. When overtaking, the overtaken vehicles (the green and blue vehicles) will give way to the overtaking vehicle (the red vehicle), which could improve the safety and efficiency of the traffic flow. After the overtaking task, every vehicle returns to its original reference path.

\begin{figure}[t]%[ht]
\centering
\includegraphics[scale=0.38]{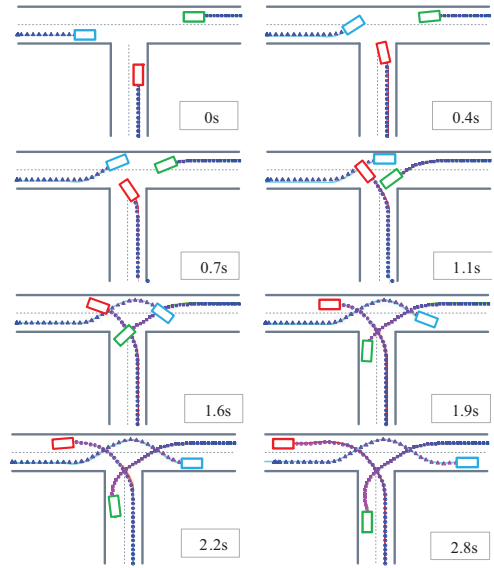}
\caption{Snapshots of coordination in the intersection of three vehicles.}
\label{figurelabel8}
\vspace{-2mm}
\end{figure}

Our second scenario is vehicle motion control at intersection without traffic lights. As shown in Fig.~\ref{figurelabel8}, the vehicles will deviate from the reference trajectory to keep a safe distance from surrounding vehicles when the conflicting may happen, \emph{i.e.}, the blue vehicle will change lane to avoid collision with the red vehicle and the green vehicle turns left in advance to keep a safe distance with the blue vehicle. This scenario tests the effectiveness of the proposed algorithm in a more complicated traffic condition, and the results are promising and meaningful for improving traffic efficiency.

The authors in \cite{gerdts2018optimization} adopted the priority approach to deal with the scenario, where a similar effect was obtained. As mentioned in the previous section, one major advantage of our algorithm is the ability of parallel computation, which is able to scale to large-scale instances. We demonstrate this fact in the next subsection.

\begin{figure}[t]%[ht]
\centering
\includegraphics[scale=0.36]{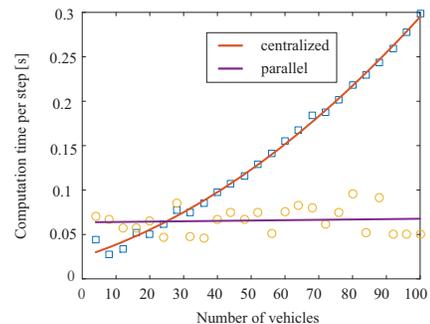}
\caption{Computation performance of the centralized and parallel algorithms: the proposed algorithm outperform the centralized one for large scale instances.}
\label{figurelabel9}
\vspace{-4mm}
\end{figure}

%%%%%%%%%%%%%%%%%%%%%%%%%%%%%%%%%%%%%%%%%%%%%%%%%%%%%%%%%%%%%%%%%%%%%%%%%%%%%%%%%%%

\subsection{Discussion of the Computation Performance}

In the discretized setting, the centralized optimal control problem amounts to solve a quadratic programming (QP) problem, which can be solved in polynomial time. In our problem, the dimension of the QP program is $N\times N_p$, where $N$ is the number of vehicles. The required number of iterations is $k = \mathcal{O}(\frac{1}{\epsilon})$ for ADMM, where $\epsilon$ is the prescribed precision. In principal, the computation complexity of the centralized and parallel solutions are $\mathcal{O}\big((N\times N_p)^m\big)$ and $\mathcal{O}\big(k(N_p)^m\big)$ respectively, which indicates the parallel algorithm is scalable to large-scale problems.

Here, we conducted numerical experiments to verify the performance of our  algorithm with different vehicle size, ranging from 4 to 100.  Fig.~\ref{figurelabel9} shows the computational time in one sampling step. The time used for communication is negligible and the accumulation of the maximum time consumption for each computation node in every iteration is used as the computation for the parallel algorithm. As the number of vehicles increases, the optimization time on each computation node is almost invariability, while the computation time for the centralized optimization increases. This is in accordance to our expectation, since the computation in the ADMM algorithm can be carried out in a parallel fashion, where the dimension of each subproblem is independent to the network size. When the vehicle number is small, the centralized optimization algorithm seems faster in our simulations. However, the computation time of the centralized method grows quickly as the number of vehicles increases, thus the centralized algorithm is not suitable for large-scale instances. In contrast, the parallel algorithm is scalable to large-scale cases.

%%%%%%%%%%%%%%%%%%%%%%%%%%%%%%%%%%%%%%%%%%%%%%%%%%%%%%%%%%%%%%%%%%%%%%%%%%%%%%%%%%%%%%%

\balance
\section{Conclusions}

This paper has proposed a parallel optimization algorithm using ADMM, which is promising for cooperative automation of large-scale connected vehicles. In particular, we formulated the cooperative automation task as a centralized optimization problem in a receding horizon fashion. %To ease the formulation of the ADMM algorithm, 
Taylor expansion was applied to the centralized problem?% based on the properties of receding horizon optimization. 
The resulting algorithm is suitable for parallel implementations. Numerical experiments based on typical traffic scenarios demonstrated the effectiveness and efficiency of the proposed algorithm.
One future work is to address the convergence of the parallel algorithm under time delay of information broadcasting. Also, validations on real world experiments would be interesting to test the performance of our algorithm.

\addtolength{\textheight}{-12cm}   % This command serves to balance the column lengths
                                  % on the last page of the document manually. It shortens
                                  % the textheight of the last page by a suitable amount.
                                  % This command does not take effect until the next page
                                  % so it should come on the page before the last. Make
                                  % sure that you do not shorten the textheight too much.

%%%%%%%%%%%%%%%%%%%%%%%%%%%%%%%%%%%%%%%%%%%%%%%%%%%%%%%%%%%%%%%%%%%%%%%%%%%%%%%%

%%%%%%%%%%%%%%%%%%%%%%%%%%%%%%%%%%%%%%%%%%%%%%%%%%%%%%%%%%%%%%%%%%%%%%%%%%%%%%%%

%%%%%%%%%%%%%%%%%%%%%%%%%%%%%%%%%%%%%%%%%%%%%%%%%%%%%%%%%%%%%%%%%%%%%%%%%%%%%%%%

\section*{Acknowledgement}

The authors would like to thank Prof. Guangyu Tian from Tsinghua University for his valuable comments on this work.

%%%%%%%%%%%%%%%%%%%%%%%%%%%%%%%%%%%%%%%%%%%%%%%%%%%%%%%%%%%%%%%%%%%%%%%%%%%%%%%%

\bibliographystyle{ieeetr}
\bibliography{sample}

\begin{thebibliography}{10}

\bibitem{6823640}
N.~Lu, N.~Cheng, N.~Zhang, X.~Shen, and J.~W. Mark, ``Connected vehicles:
  Solutions and challenges,'' {\em IEEE Internet Things J.}, vol.~1, no.~4,
  pp.~289--299, 2014.

\bibitem{6121907}
J.~Lee and B.~Park, ``Development and evaluation of a cooperative vehicle
  intersection control algorithm under the connected vehicles environment,''
  {\em IEEE Trans. Intell. Transp. Syst.}, vol.~13, pp.~81--90, March 2012.

\bibitem{li2017dynamical}
S.~E. Li, Y.~Zheng, K.~Li, Y.~Wu, J.~K. Hedrick, F.~Gao, and H.~Zhang,
  ``Dynamical modeling and distributed control of connected and automated
  vehicles: Challenges and opportunities,'' {\em IEEE Intell. Transp. Syst.
  Mag.}, vol.~9, no.~3, pp.~46--58, 2017.

\bibitem{li2017simultaneous}
B.~Li, Y.~Zhang, Z.~Shao, and N.~Jia, ``Simultaneous versus joint computing: A
  case study of multi-vehicle parking motion planning,'' {\em J. Comput. Sci.},
  vol.~20, pp.~30--40, 2017.

\bibitem{kuwata2011cooperative}
Y.~Kuwata and J.~P. How, ``Cooperative distributed robust trajectory
  optimization using receding horizon milp,'' {\em IEEE Trans. Control Syst.
  Technol}, vol.~19, no.~2, pp.~423--431, 2011.

\bibitem{zheng2017distributed}
Y.~Zheng, S.~E. Li, K.~Li, F.~Borrelli, and J.~K. Hedrick, ``Distributed model
  predictive control for heterogeneous vehicle platoons under unidirectional
  topologies,'' {\em IEEE Trans. Control Syst. Technol}, vol.~25, no.~3,
  pp.~899--910, 2017.

\bibitem{wu2016distributed}
Y.~Wu, S.~E. Li, Y.~Zheng, and J.~K. Hedrick, ``Distributed sliding mode
  control for multi-vehicle systems with positive definite topologies,'' in
  {\em IEEE Conf. on Decision and Control}, pp.~5213--5219, 2016.

\bibitem{velagapudi2010decentralized}
P.~Velagapudi, K.~Sycara, and P.~Scerri, ``Decentralized prioritized planning
  in large multirobot teams,'' in {\em IEEE Conf. Intell. Robot. and Syst.},
  pp.~4603--4609, 2010.

\bibitem{plessen2016multi}
M.~G. Plessen, D.~Bernardini, H.~Esen, and A.~Bemporad, ``Multi-automated
  vehicle coordination using decoupled prioritized path planning for multi-lane
  one-and bi-directional traffic flow control,'' in {\em IEEE Conf. on Decision
  and Control}, pp.~1582--1588, 2016.

\bibitem{yu2016optimal}
J.~Yu and S.~M. LaValle, ``Optimal multirobot path planning on graphs: Complete
  algorithms and effective heuristics,'' {\em IEEE Trans. Robot.}, vol.~32,
  no.~5, pp.~1163--1177, 2016.

\bibitem{zheng2016stability}
Y.~Zheng, S.~E. Li, J.~Wang, D.~Cao, and K.~Li, ``Stability and scalability of
  homogeneous vehicular platoon: Study on the influence of information flow
  topologies,'' {\em IEEE Trans. Intell. Transp. Syst.}, vol.~17, no.~1,
  pp.~14--26, 2016.

\bibitem{richards2002aircraft}
A.~Richards and J.~P. How, ``Aircraft trajectory planning with collision
  avoidance using mixed integer linear programming,'' in {\em Proc. Amer.
  Control Conf.}, vol.~3, pp.~1936--1941, 2002.

\bibitem{borrelli2006comparison}
F.~Borrelli, D.~Subramanian, A.~U. Raghunathan, and L.~T. Biegler, ``A
  comparison between milp and nlp techniques for centralized trajectory
  planning of multiple unmanned air vehicles,'' in {\em Proc. Amer. Control
  Conf.}, 2006.

\bibitem{zheng2017chordal}
Y.~Zheng, G.~Fantuzzi, A.~Papachristodoulou, P.~Goulart, and A.~Wynn, ``Chordal
  decomposition in operator-splitting methods for sparse semidefinite
  programs,'' {\em arXiv preprint arXiv:1707.05058}, 2017.

\bibitem{alonso2017multi}
J.~Alonso-Mora, S.~Baker, and D.~Rus, ``Multi-robot formation control and
  object transport in dynamic environments via constrained optimization,'' {\em
  Int. J. Robot. Res.}, vol.~36, no.~9, pp.~1000--1021, 2017.

\bibitem{urcola2017cooperative}
P.~Urcola, M.~T. L{\'a}zaro, J.~A. Castellanos, and L.~Montano, ``Cooperative
  minimum expected length planning for robot formations in stochastic maps,''
  {\em Rob. and Auton. Syst.}, vol.~87, pp.~38--50, 2017.

\bibitem{zheng2018scalable}
Y.~Zheng, R.~P. Mason, and A.~Papachristodoulou, ``Scalable design of
  structured controllers using chordal decomposition,'' {\em IEEE Trans. Autom.
  Control}, vol.~63, no.~3, pp.~752--767, 2018.

\bibitem{boyd2011distributed}
S.~Boyd, N.~Parikh, E.~Chu, B.~Peleato, J.~Eckstein, {\em et~al.},
  ``Distributed optimization and statistical learning via the alternating
  direction method of multipliers,'' {\em Found. Trends Mach. Learn.}, vol.~3,
  no.~1, pp.~1--122, 2011.

\bibitem{bento2013message}
J.~Bento, N.~Derbinsky, J.~Alonso-Mora, and J.~S. Yedidia, ``A message-passing
  algorithm for multi-agent trajectory planning,'' in {\em Adv. Neural Inf.
  Process Syst.}, pp.~521--529, 2013.

\bibitem{paden2016survey}
B.~Paden, M.~{\v{C}}{\'a}p, S.~Z. Yong, D.~Yershov, and E.~Frazzoli, ``A survey
  of motion planning and control techniques for self-driving urban vehicles,''
  {\em IEEE Trans. Intell. Vehi.}, vol.~1, no.~1, pp.~33--55, 2016.

\bibitem{ong2015cooperative}
H.~Y. Ong and J.~C. Gerdes, ``Cooperative collision avoidance via proximal
  message passing,'' in {\em Proc. Amer. Control Conf.}, pp.~4124--4130, 2015.

\bibitem{erseghe2014distributed}
T.~Erseghe, ``Distributed optimal power flow using admm,'' {\em IEEE Trans.
  Power Syst.}, vol.~29, no.~5, pp.~2370--2380, 2014.

\bibitem{zheng2017fast}
Y.~Zheng, G.~Fantuzzi, A.~Papachristodoulou, P.~Goulart, and A.~Wynn, ``Fast
  {ADMM} for semidefinite programs with chordal sparsity,'' in {\em Proc. Amer.
  Control Conf.}, pp.~3335--3340, 2017.

\bibitem{zheng2017fastadmm}
H.~Zheng, R.~R. Negenborn, and G.~Lodewijks, ``Fast {ADMM} for distributed
  model predictive control of cooperative waterborne agvs,'' {\em IEEE Trans.
  Control Syst. Technol.}, vol.~25, no.~4, pp.~1406--1413, 2017.

\bibitem{optimization2017inc}
G.~Optimization, ``Inc.,“gurobi optimizer reference manual,” 2017,'' {\em
  URL: http://www. gurobi. com}, 2017.

\bibitem{gerdts2018optimization}
M.~Gerdts and B.~Martens, ``Optimization-based motion planning in virtual
  driving scenarios with application to communicating autonomous vehicles,''
  {\em arXiv preprint arXiv:1801.07612}, 2018.

\end{thebibliography}

\end{document}